\documentstyle[prb,preprint,aps]{revtex}
\draft
\input{psfig}
\psfigurepath{FIG/}

\def\AmS{{\protect\the\textfont2
        A\kern-.1667em\lower.5ex\hbox{M}\kern-.125emS}}

\makeatletter
\def\thepage{1-\@arabic\c@page}
\def\@pnumwidth{2em}
\makeatother
\begin{document}

\title{Bandgap Change of Carbon Nanotubes: Effect of Small Uniaxial
and Torsional Strain}
\author{Liu Yang$^1$, M. P. Anantram$^2$~\cite{byline2},
Jie Han$^2$ and  J. P. Lu$^2$~\cite{byline4} \\}
\address{$^1$ Thermosciences Institute, NASA Ames Research Center,
            Mail Stop 230-3, Moffett Field, CA, USA 94035-1000}
\address{$^2$ NASA Ames Research Center, Mail Stop T27A-1,
           Moffett Field, CA, USA 94035-1000 }

\maketitle

\begin{abstract}
We use a simple picture based on the $\pi$ electron approximation to
study the bandgap variation of carbon nanotubes with uniaxial and 
torsional strain. We find (i) that the 
magnitude of slope of bandgap versus strain has an almost universal
behaviour that depends on the chiral angle, (ii) that the sign of slope
depends on the value of $(n-m) \bmod 3$ and (iii) a novel change in
sign of the slope of bandgap versus uniaxial strain arising from a
change in the value of the quantum number corresponding to the minimum
bandgap. Four orbital calculations are also presented to show that the
$\pi$ orbital results are valid.
\end{abstract}


\section{Introduction}

The mechanical and electronic properties of carbon nanotubes (CNT)
have individually been studied in some
detail~\cite{Mintmire92,Hamada92,Saito92,Dresselhaus_book,Yakobson}
and the predicted dependence of bandgap on 
chirality~\cite{Mintmire92,Hamada92,Saito92}
has been observed.~\cite{Wildoer98} The study of bandgap variation
with mechanical deformation is important in view of the ability to
manipulate individual nanotubes~\cite{Venema97}.
Additionally, they could form the basis for nanoscale sensors in a
manner similar to experiments using $C_{60}$ 
molecules.~\cite{Joachim95} Recent studies of bandgap change of zigzag
and armchair tubes on mechanical strain have shown interesting 
behavior.~\cite{Heyd97,Brenner_unpublished,Kane97} Refs. 
\onlinecite{Heyd97} and \onlinecite{Brenner_unpublished}
studied the effect of uniaxial strain using a Green's function method
based on the $\pi$ electron approximation and a four orbital numerical
method, respectively. Ref. \onlinecite{Kane97} predicted the opening 
of a bandgap in armchair tubes under torsion, using a method that wraps
a massless two dimensional Dirac Hamiltonian on a curved surface. In 
this paper, we present a simple and unified picture of the band gap 
variation of chiral and achiral CNT with uniaxial and torsional strain.
The method used is discussed in section \ref{section:method}.
The results obtained by using a single $\pi$ orbital are discussed in
section \ref{subsect:pi_orbital} and are compared to four orbital
calculations in section \ref{subsect:4_orbital}.
The conclusions are presented in section \ref{section:conclusions}.

\section{Method}
\label{section:method}

In the presence of a uniform uniaxial and torsional strain, a distorted
graphene sheet continues to have two atoms per unit cell [Fig. 1]. It
is convenient to represent the change in bond lengths using the chirality
dependent coordinate system.
The axes of the chirality dependent coordinate system corresponding to 
$(n,m)$ CNT are the line joining the $(0,0)$ and $(n,m)$ carbon 
atoms ($\hat{c}$), and its perpendicular ($\hat{t}$)
[Fig. 1].~\cite{coord_sys} The fixed and chirality dependent coordinate
system are related by,
$\hat{c} = cos \theta \hat{x} + sin \theta \hat{y}$ and
$\hat{t} = -sin \theta \hat{x} + cos \theta \hat{y}$,where,
$sin(\theta) = \frac{1}{2} \frac{n-m}{c_h}$ and
$cos(\theta) = \frac{\sqrt{3}}{2} \frac{n+m}{c_h}$.
$c_h = \sqrt{n^2 + m^2 + n m}$, is the circumference of the tube in
units of the equilibrium lattice vector length,
$|\vec{a}_1| = |\vec{a}_2| = a_0$.
The bond vectors are given by,
\begin{eqnarray}
\vec{r}_1 = \frac{a_0}{2} \frac{n+m}{c_h} \hat{c} - \frac{a_0}{2 \sqrt{3}}
\frac{n-m}{c_h} \hat{t} + \delta \vec{r}_1 \mbox{ and }
\vec{r}_2 = -\frac{a_0}{2} \frac{m}{c_h} \hat{c} + \frac{a_0}{2 \sqrt{3}}
\frac{2n+m}{c_h} \hat{t} + \delta \vec{r}_2  \mbox{ ,} 
\label{eq:r_1}
\end{eqnarray}
where, $\delta \vec{r}_i$ represents deviation from an undistorted
sheet and $\vec{r}_3 = - ( \vec{r}_1 + \vec{r}_2 )$. Within the 
context of continuum mechanics, application of a uniaxial or torsional
strain causes the following change in the bond vectors of Fig. 1:
\begin{eqnarray}
r_{i t} \rightarrow (1+\epsilon_t) r_{i t} \;
\; \mbox{and} \;\; r_{i c} \rightarrow (1+\epsilon_c) r_{i c} \;\;
                     \mbox{(tensile)} \label{eq:coord_tensile} \\
r_{i c} \rightarrow r_{i c} + tan(\gamma) r_{i t} \;\;
\mbox{(torsion)} \mbox{,} \label{eq:coord_torsion}
\end{eqnarray}
where, $i = 1,2,3$ and $r_{ip}$ is the p-component of $\vec{r_i}$
($p = {c},{t}$).
$\epsilon_t$ and $\epsilon_c$ represent the strain along $\hat{t}$ and
 $\hat{c}$ respectively, in the case of uniaxial strain.
$\gamma$ is the shear strain.

Using Eqs. (\ref{eq:r_1})-(\ref{eq:coord_torsion}), the lattice
vectors of the distorted sheet are,
\begin{eqnarray}
\vec{a}_1 &=& \vec{r}_1 - \vec{r}_3 = a_0 [ (1+\epsilon_c) \frac{1}{2}
\frac{2n+m}{c_h} + tan(\gamma) \frac{\sqrt{3}}{2} \frac{m}{c_h} ] \hat{c}
+ a_0 (1+\epsilon_t) \frac{\sqrt{3}}{2} \frac{m}{c_h} \hat{t}
\label{eq:a1} \\
\vec{a}_2 &=& \vec{r}_1 - \vec{r}_2 = a_0 [ (1+\epsilon_c) \frac{1}{2}
\frac{n+2m}{c_h} - tan(\gamma) \frac{\sqrt{3}}{2} \frac{n}{c_h} ] \hat{c}
- a_0 (1+\epsilon_t) \frac{\sqrt{3}}{2} \frac{n}{c_h} \hat{t}
\label{eq:a2} \mbox{ .}
\end{eqnarray}
The corresponding unit cell area is
$|\vec{a}_1 \times \vec{a}_2|  = \frac{\sqrt{3}}{2} (1+\epsilon_t)
(1+\epsilon_c) a_0^2$.
The real space unit cells correspond to $\vec{r}= j_1 \vec{a}_1
+ j_2 \vec{a}_2$, where $j_1$ and $j_2$ are integers. The 1D unit cell
length ($T$) is the shortest $r_{t}$ for which $r_c=0$.
That is, the two lattice points, $\vec{r} = 0$ and $\vec{r} = j_1 
\vec{a}_1 + j_2 \vec{a}_2$ have the same $\hat{c}$ coordinate. This
corresponds to the following condition on $j_i$ and $j_2$,
\begin{eqnarray}
(1+\epsilon_c) [ j_1 (2n+m) + j_2 (n+2m) ] + 
tan(\gamma) \sqrt{3} [j_1 m -j_2 n] = 0 \label{eq:cond}
\end{eqnarray}
and the 1D  unit cell length is,
\begin{eqnarray}
T = a_0 (1+\epsilon_t) \frac{\sqrt{3}}{2}
\frac{(j_1 m - j_2 n)}{c_h}
\label{eq:1Dunit_cell}
\end{eqnarray}

When only uniaxial strain is present ($\gamma=0$), Eq. (\ref{eq:cond})
corresponds to, $j_1 (2n+m) + j_2 (n+2m) = 0$. 
The corresponding $j_1$ and $j_2$ with smallest absolute values are
$j_1 = (n+2m)/gcd(2n+m,n+2m)$ and $j_2 = - (2n+m)/gcd(2n+m,n+2m)$.
$gcd$ refers to the greatest common divisor.
Using these values in Eq.(\ref{eq:1Dunit_cell}), the 1D unit
cell length of an (n,m) tube is,
\begin{eqnarray}
T = (1+\epsilon_t) \sqrt{3} c_h a_0 / gcd(2n+m,n+2m) \mbox{ .}
\label{eq:1Dunit_cell_tensile}
\end{eqnarray}
In the absence of strain, Eq. (\ref{eq:1Dunit_cell_tensile}) reduces 
to the result for undeformed nanotubes. In the presence of uniaxial strain,
the unit cell length is equal to $(1+\epsilon_t)$ times the unstrained
unit cell length. When only torsion is present, Eq. (\ref{eq:cond}) 
simplifies to,
\begin{eqnarray}
j_1 (2n+m) + j_2 (n+2m) + tan(\gamma) \sqrt{3} (j_1 m -j_2 n) = 0
					\label{eq:cond_torsion}
\mbox{ .}
\end{eqnarray}

For arbitrary values of $\gamma$, $n$ and $m$, this equation
corresponds to a large $T$. For example, from Fig. 1 it is easy to see
that under torsion, the unit cell of an armchair tube can be much
larger than $a_0$ depending on the value of $\gamma$. We will come
back to this point at the end of section \ref{section:method}, where
we discuss calculation of bandgap change due to torsion.

We treat the nanotube within the approximation that it is a rolled up
graphene sheet and assume a single $\pi$ orbital per carbon atom. We
calculate the band structure of the distorted sheet to 
be,~\cite{Wallace47}
\begin{eqnarray}
E(\vec{k}) = ( t_1^2 + t_2^2 + t_3^2
      + 2 t_1 t_2 \; \mbox{cos} [\vec{k}\cdot(\vec{r}_1-\vec{r}_2)]
      + 2 t_2 t_3 \; \mbox{cos} [\vec{k}\cdot(\vec{r}_2-\vec{r}_3)]
      + 2 t_3 t_1 \; \mbox{cos} [\vec{k}\cdot(\vec{r}_3-\vec{r}_1)]
)^\frac{1}{2} \mbox{ ,} \label{eq:band1}
\end{eqnarray}
where, $\vec{k} = k_c \hat{c} + k_t \hat{t}$.
The primary effects of change in bond vectors are to alter the hopping
parameter between carbon atoms and lattice vectors. 
The hopping parameter is assumed to scale with bond length
as,~\cite{Harrison} $t_i = t_0\;(r_0/r_i)^2$, where $t_0$ and $r_0$
are the hopping parameter and bond length of an unstrained graphene 
sheet. 
The value of $t_0$ is around 3eV.
From Eqs. (\ref{eq:a1}) and (\ref{eq:a2}), the circumference 
of the distorted sheet is $(1+\epsilon_c) c_h a_0$. The wave function of
the CNT is quantized around the circumference and so $k_c$ is given by,
\begin{eqnarray}
k_c (1+\epsilon_c) c_h a_0 = 2\pi q \mbox{ ,} \label{eq:q}
\end{eqnarray}
where, $q$ is an integer. Eq. (\ref{eq:band1}) can now be
written as,
\begin{eqnarray}
E(k_t) = ( t_1^2 + t_2^2 + t_3^2 
            &+& 2 t_1 t_2 \; \mbox{cos} [\pi q \frac{n+2m}{c_h^2}
- \frac{\sqrt{3}}{2} \frac{n}{c_h} k_t^\prime a_0 
- \; \pi q \frac{\sqrt{3} \; tan(\gamma)}{1+\epsilon_c} \frac{n}{c_h^2}\; ]
                                                       \nonumber\\
           &+& 2 t_1 t_3 \; \mbox{cos} [\pi q \frac{2n+m}{c_h^2} 
+ \frac{\sqrt{3}}{2} \frac{m}{c_h} k_t^\prime a_0 
+ \; \pi q \frac{\sqrt{3} \; tan(\gamma)}{1+\epsilon_c} \frac{m}{c_h^2}\; ]
                                                       \nonumber\\
           &+&  2 t_2 t_3 \; \mbox{cos} [\pi q \frac{n-m}{c_h^2} 
+ \frac{\sqrt{3}}{2} \frac{n+m}{c_h} k_t^\prime a_0
+ \; \pi q \frac{\sqrt{3} \; tan(\gamma)}{1+\epsilon_c} \frac{n+m}{c_h^2}\; ]
)^\frac{1}{2} \mbox{ ,} \label{eq:band}
\end{eqnarray}
where, $k_t^\prime = (1+\epsilon_t) k_t$.
The bandgap of an $(n,m)$ tube in presence of uniaxial 
($\gamma=0$) or torsional strain ($\epsilon_c = \epsilon_t = 0$) can
be easily calculated from Eq. (\ref{eq:band}). In case of uniaxial
strain, the limits of $k_t$ are given by 
$-\frac{\pi}{T} < k_t < \frac{\pi}{T}$, where $T$ is the
1D lattice vector length determined by 
Eq. (\ref{eq:1Dunit_cell_tensile}). 
The number of atoms in the 1D unit cell does not change in the presence
of uniaxial strain and so the
range of $q$ does not change from the undeformed case ($q=0, 1, 2, ...,
N_c$, where $N_c$ is the number of hexagons in the 1D unit cell).

In the case of torsion, the number of atoms in the 1D unit cell and $T$
can be large [Eq. (\ref{eq:cond_torsion})]. The corresponding span of
$k_t$ is then small compared to the undeformed tube and the range of 
$q$ is commensurate with the number of atoms in the 1D unit cell. The
eigen spectrum can however be obtained from Eq. (\ref{eq:band})
by setting $\gamma=0$ and
spanning over the same values of $q$ and $k_t$ as in the undeformed
case. This is because the eigen spectrum depends only on the tight
binding parameters (and not on the exact geometry) if the coordination
number of the carbon atoms remains constant.~\cite{footnote3}

\section{Results and Discussion}
\label{section:results}

The results obtained using the method described in section
\ref{section:method} are discussed in section \ref{subsect:pi_orbital}.
We then present the results from four orbital calculations with energy
minimized structures in section \ref{subsect:4_orbital}.

\subsection{$\pi$ orbital}
\label{subsect:pi_orbital}

We first consider the case of uniaxial strain.
The bandgap is obtained by finding the minimum of $E(k_t)$, where the
span of $k_t$ and $q$ are discussed below Eq. (\ref{eq:band}). 
The bandgap change is largest for zigzag tubes and the magnitude of
$|dE_g/d\sigma|$ is approximately equal to $3t_0$. For armchair tubes,
application of uniaxial strain does not cause a bandgap. We find that,
(i) $|dE_g/d\sigma|$ increases with increase in chiral angle [Fig. 2]
and (ii) the sign of $dE_g/d\sigma$ follows the $(n-m) \bmod 3$
rule.~\cite{footnote1} For example, the chiral angle of $(6,5)$ and
$(6,4)$ tubes are close to that of armchair tubes. The slope of bandgap
versus strain is correspondingly small and the sign of slope are
opposite.  For semiconducting zigzag tubes and armchair tubes,
our results agree with Ref. \onlinecite{Heyd97}.

As uniaxial strain increases, there is an abrupt reversal in sign of
${dE_g}/{d\sigma}$ as illustrated for zigzag tubes in Fig. 4.
This feature indicates a change in band index $q$ corresponding to the
bandgap and can be understood from the following expression
that describes dependence of energy for various values of $q$ at $k_t=0$
[Eq. (\ref{eq:slope1}) of appendix]:
\begin{eqnarray}
E(0) = E_0(q) \; - \; 2t_0  \; \frac{\delta r_1}{r_0}
 \; \left[ 1 - \frac{2\delta r_2}{\delta r_1} \;
                  \mbox{cos}(\frac{q\pi}{n}) \right ] \; sgn(x)
\mbox{ ,} \label{eq:slope}
\end{eqnarray}
where, $sgn(x) = \left[ 1 - 2\mbox{cos}({q\pi}/{n}) \right]$. The
minimum value of $E_{0}(q) = t_0 \; |1 - 2\mbox{cos}({q\pi}/{n})|$ is
half of the bandgap of an unstrained tube. The first term of
Eq. (\ref{eq:slope}) takes the smallest value for the band index
$q=q_0$ that satisfies $n=3q_0\pm1$. The second term can however change
sign when $q$ changes from $q_0$ to $q_0\pm1$. As a result, a
dramatic change in sign of ${dE_g}/{d\sigma}$ becomes possible if
the magnitude of the second term is larger than change in the first
term [Fig. 3]. The strain required to observe this effect decreases
as the inverse radius of tube for large $n$. This is because the 
difference in energy of
the $q_0$ and $q_0 \pm 1$ bands become smaller with increase
in radius. Fig. 3 demonstrates this point by comparing the (10,0) and 
(19,0) tubes. For the (19,0) tube, the change in slope occurs at around 
five percent strain. These strain values are accessible in bulk 
nanotube samples.~\cite{aps99} The inset of Fig. 3 shows change in 
energy of the q=6 and q=7 bands for the (19,0) tube for three different 
values of strain. While the q=6 band shifts up in energy as strain 
increases, the q=7 band shifts down. Thus leading to the discussed 
change in sign of ${dE_g}/{d\sigma}$.

In case of torsional strain, the bandgap is obtained by finding the
minimum of $E(k_t)$ using Eq. (\ref{eq:band}), where the span of $k_t$ and
$q$ are discussed in the last paragraph of section
\ref{section:method}. The magnitude of $|dE_g/d\sigma|$ is 
approximately equal to $3t_0$ for armchair tubes and this is in
agreement with Ref. \onlinecite{Kane97}. For 
zigzag tubes, torsion causes only a small change in bandgap. The 
leading term in bandgap change depends on $\gamma$ only to second
order. We find that, (i) $|dE_g/d\sigma|$ decreases
with increase in chiral angle and takes the smallest value for zigzag
tubes  [Fig. 4] and (ii) the sign of $dE_g/d\sigma$ follows the
$(n-m) \bmod 3$ rule.~\cite{footnote1}

\subsection{Four orbital}
\label{subsect:4_orbital}

To verify the simple picture presented, we have also performed four
orbital calculations using the parametrization given in Ref.
\onlinecite{Tomanek88}.
The change in bond lengths are computed using both continuum mechanics
[Eqs. (\ref{eq:coord_tensile}) and (\ref{eq:coord_torsion})] and
structures that are energy minimized by Brenner
potential.~\cite{Brenner90}
The energy minimization was performed with periodic boundary conditions.
For the small values of strain considered, we find that the bandgap is
not very sensitive to the two methods of obtaining the bond lengths.
The results presented in Figs. 5 and 6 correspond to the bond lengths
obtained by energy minimization.
For semiconducting tubes, the results of Figs. 5 and 6 agree with
the $\pi$ orbital results presented in Figs. 2 and 4 respectively:
The slope of $dE_g/d\sigma$ follows the $(n-m) \bmod 3$ rule and
the magnitude of slope varies monotonically with chiral angle.
The primary difference concerns non armchair tubes satisfying $n-m =
3*integer$.
This is not surprising because Ref. \onlinecite{Hamada92} has predicted
such tubes to have a small bandgap due to curvature induced
hybridization at zero strain.
As a result, applying either tension or compression does not produce
the "V" shaped curve of Fig. 2 with zero bandgap at zero strain.
The difference is that the curves are shifted away from the origin
as shown in Fig. 5.

\section{Conclusions}
\label{section:conclusions}

In conclusion, we present a simple picture to calculate bandgap
versus strain of CNT with arbitrary chirality. We find that under 
uniaxial strain, $|dE_g/d\sigma|$ of zigzag tubes is $3t_0$ independent
of diameter, and continually decreases as the chirality changes to 
armchair, when it takes the value zero. In contrast, we show that under
torsional strain, $|dE_g/d\sigma|$ of armchair tubes is $3t_0$
independent of diameter, and continually decreases as the chirality 
changes to zigzag, where is takes a small value. The sign of
$dE_g/d\sigma$ follows the $(n-m) \bmod 3$ rule in both 
cases.~\cite{footnote1}
We also predict a change in the sign of $dE_g/d\sigma$ as a function of
strain, corresponding to a change in the value of $q$ that corresponds
to the bandgap minimum. Comparison to four orbital calculations show
that the main conclusions are unchanged. The primary difference
involves nonarmchair tubes that satisfy $n-m=3*integer$.

This work is supported by NASA contracts NAS2-14031 to Eloret (LY),
NASA Ames Research Center and U.S. Department of Energy (JPL).

\section {Appendix}

\noindent
{\it Zigzag tubes under tension:}
Under uniaxial strain the band structure of $(n,0)$ is,
\begin{eqnarray}
E(k_t) &=& \pm t_2 \left[ 1 \; \pm  \; (\frac{4t_1}{t_2}) \;
               \mbox{cos}(\frac{q\pi}{n}) \;
               \mbox{cos}((1+\epsilon_t) \frac{\sqrt{3}}{2} k_t a_0) \;
         + \; (\frac{2t_1}{t_2})^2 \; \mbox{cos}^2(\frac{q\pi}{n})
               \right]^\frac{1}{2} \mbox{ .} \label{eq:zz1}
\end{eqnarray}
$t_1 = t_3$ due to symmetry. The minimum of $E(k_t)$ occurs at $k_t=0$,
\begin{eqnarray}
E(0) = \pm \; t_2 \; | 1 - \frac{2t_1}{t_2} \;
\mbox{cos}(\frac{q\pi}{n}) | \mbox{ .}  \label{eq:Eg_zz}
\end{eqnarray}
To first order in $\delta r_i$ Eqn. (\ref{eq:Eg_zz})
is,~\cite{footnote2}
\begin{eqnarray}
E(0) = E_0(q) \; - \; 2t_0  \; \frac{\delta r_1}{r_0}
 \; \left[ 1 - \frac{2\delta r_2}{\delta r_1} \;
                  \mbox{cos}(\frac{q\pi}{n}) \right ] \; sgn(x)
\left[ 1 - 2\mbox{cos}(\frac{q\pi}{n}) \right]
\mbox{ ,} \label{eq:slope1}
\end{eqnarray}
where, $sgn(x) = \left[ 1 - 2\mbox{cos}({q\pi}/{n}) \right]$ and
$E_{0}(q) = t_0 \; |1 - 2\mbox{cos}({q\pi}/{n})|$.

\pagebreak

\noindent
{\bf Figure Captions:}

\vspace{0.2in}

\noindent
Fig. 1: The fixed $(x,y)$ and chirality dependent $(\hat{c},\hat{t})$
coordinates. $r_1$, $r_2$ and $r_3$ correspond to bonds $1$, $2$ and 
$3$ respectively. $\vec{a}_1$ and $\vec{a}_2$ are the lattice vectors
of the two dimensional sheet.

\vspace{0.2in}

\noindent
Fig. 2: Bandgap versus tensile strain:
For semiconducting tubes, the sign of slope of
$d(Bandgap)/d(Strain)$ depends only on the value of $(n - m) \bmod 3$.
The magnitude of $d(Bandgap)/d(Strain)$ is largest for zizag tubes and
decreases with decrease in chiral angle. The magnitude is smallest for
armchair tubes.
The solid, dashed and dotted
lines correspond to $(n - m) \bmod 3$ values of 1, -1 and 0 
respectively.
The value of $t_0$ is around 3eV.
\vspace{0.2in}

\noindent
Fig. 3: 
The change in slope of the (10,0) and (19,0) tubes around 10$\%$ and 
5$\%$ strain respectively is due to a change in the quantum number $q$
that yields the minimum bandgap. Inset: E vs k of the q=7 (solid) and
q=6 (dashed) bands as a function of strain for a (19,0) tube.
Strains of 0$\%$, 3$\%$ and 6$\%$ correspond to increasing thickness 
of the lines.

\vspace{0.2in}
\noindent
Fig. 4: Bandgap versus torsional strain:
For semiconducting tubes, the sign of slope of
$d(Bandgap)/d(Strain)$ depends only on the value of $(n - m) \bmod 3$.
The magnitude of $d(Bandgap)/d(Strain)$ is largest for armchair tubes
and decreases with increase in chiral angle. The magnitude is smallest
for zigzag tubes.
The solid, dashed and dotted
lines correspond to $(n - m) \bmod 3$ values of 1, -1 and 0
respectively.

\vspace{0.2in}
\noindent
Fig. 5: Same as Fig. 2 with the only difference that these are four
orbital results. In the y-axis label, t=2.66eV.

\vspace{0.2in}
\noindent
Fig. 6: Same as Fig. 4 with the only difference that these are four
orbital results. In the y-axis label, t=2.66eV.

\begin{figure}[h]
\centerline{\psfig{file=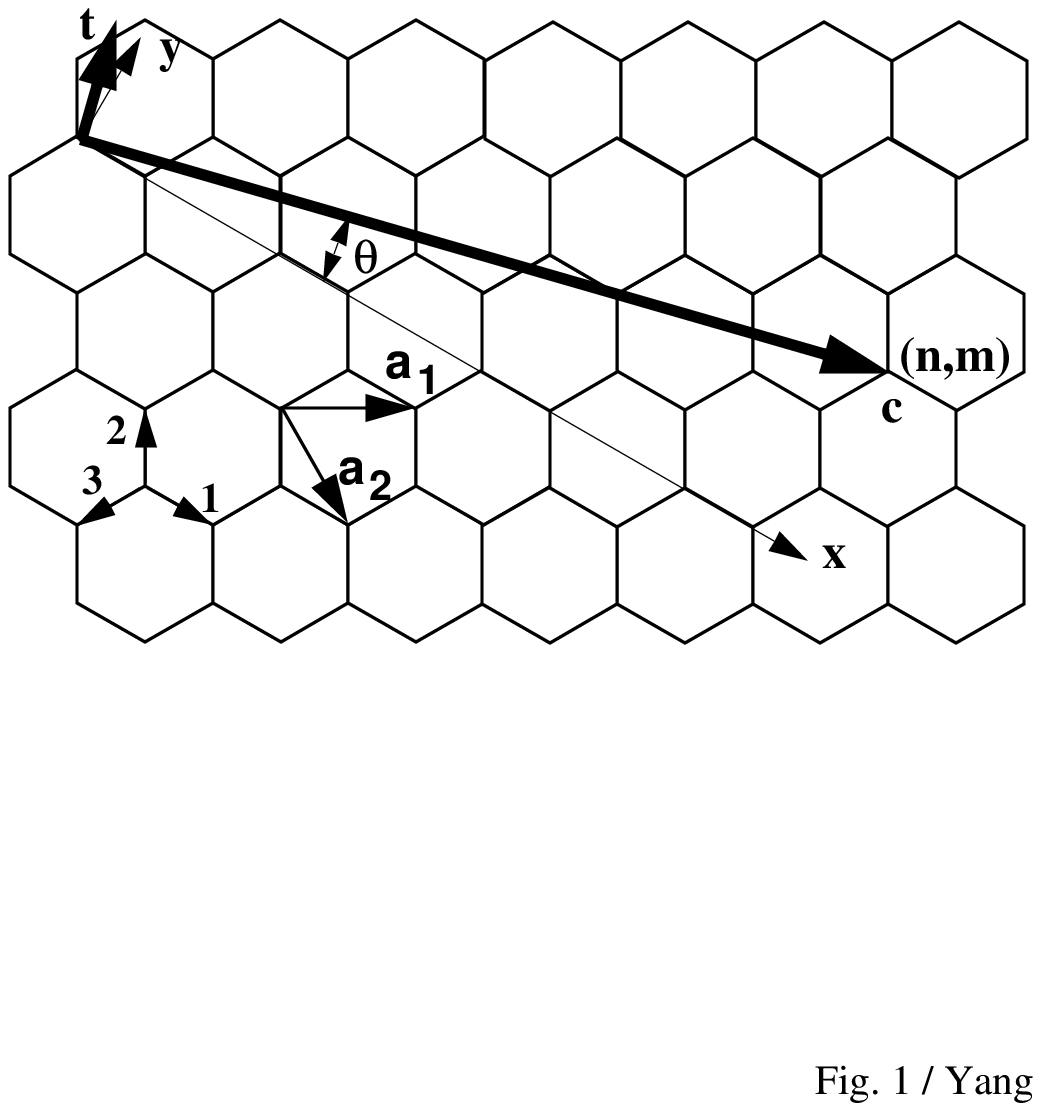,width=5in}}
\small
\end{figure}

\pagebreak

\begin{figure}[h]
\centerline{\psfig{file=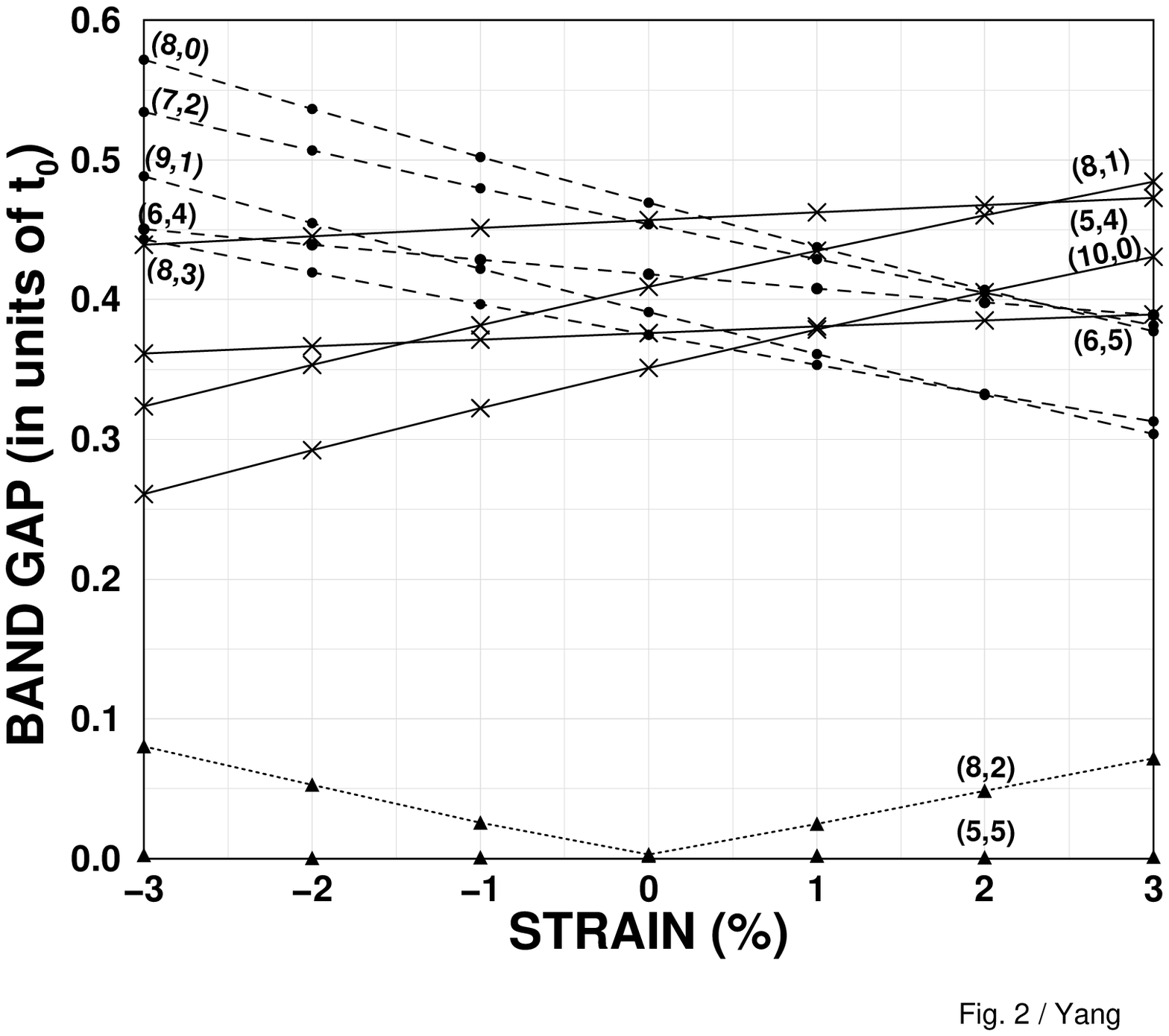,width=5in}}
\small
\end{figure}

\begin{figure}[h]
\centerline{\psfig{file=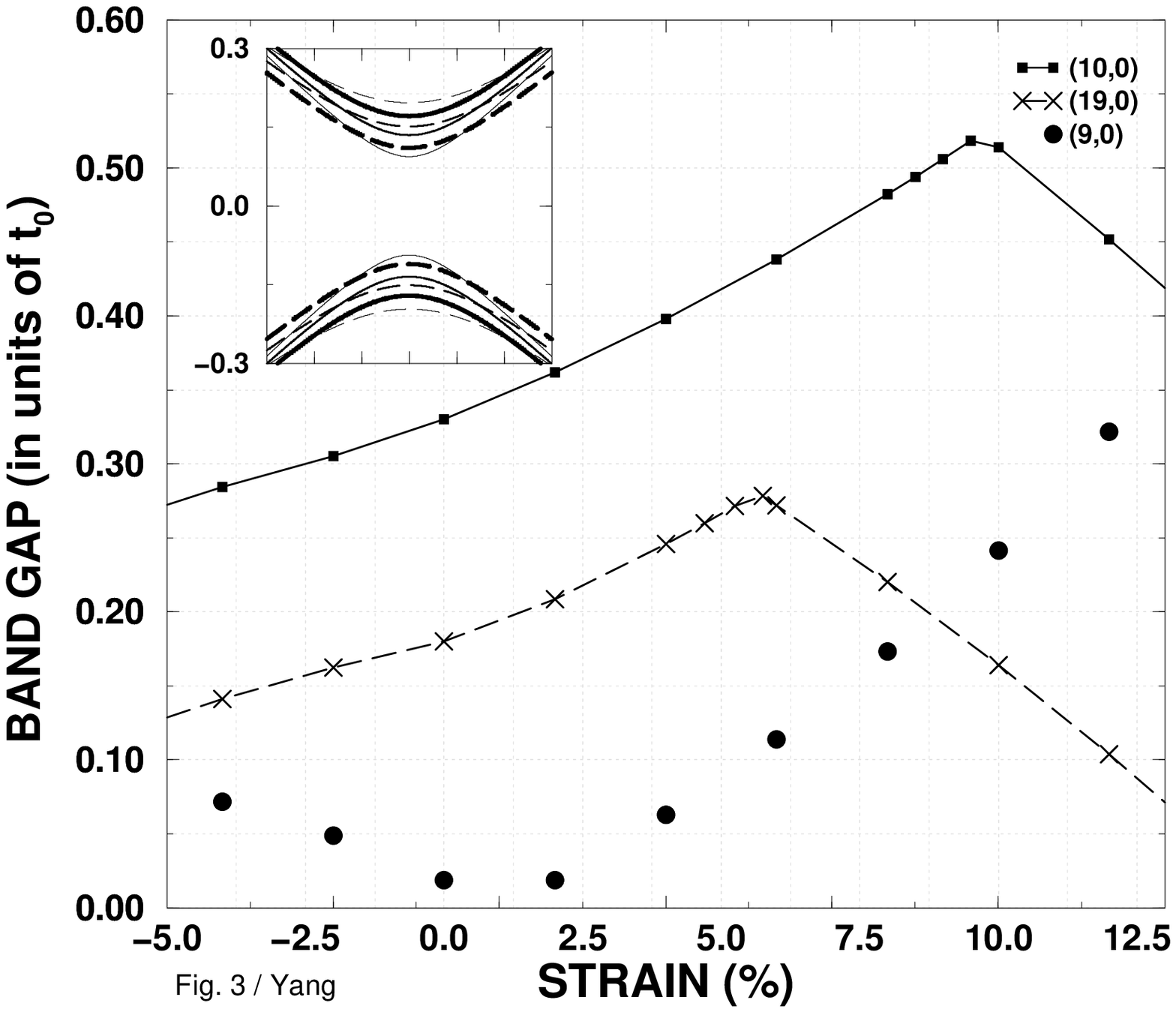,width=5in}}
\small
\end{figure}

\begin{figure}[h]
\centerline{\psfig{file=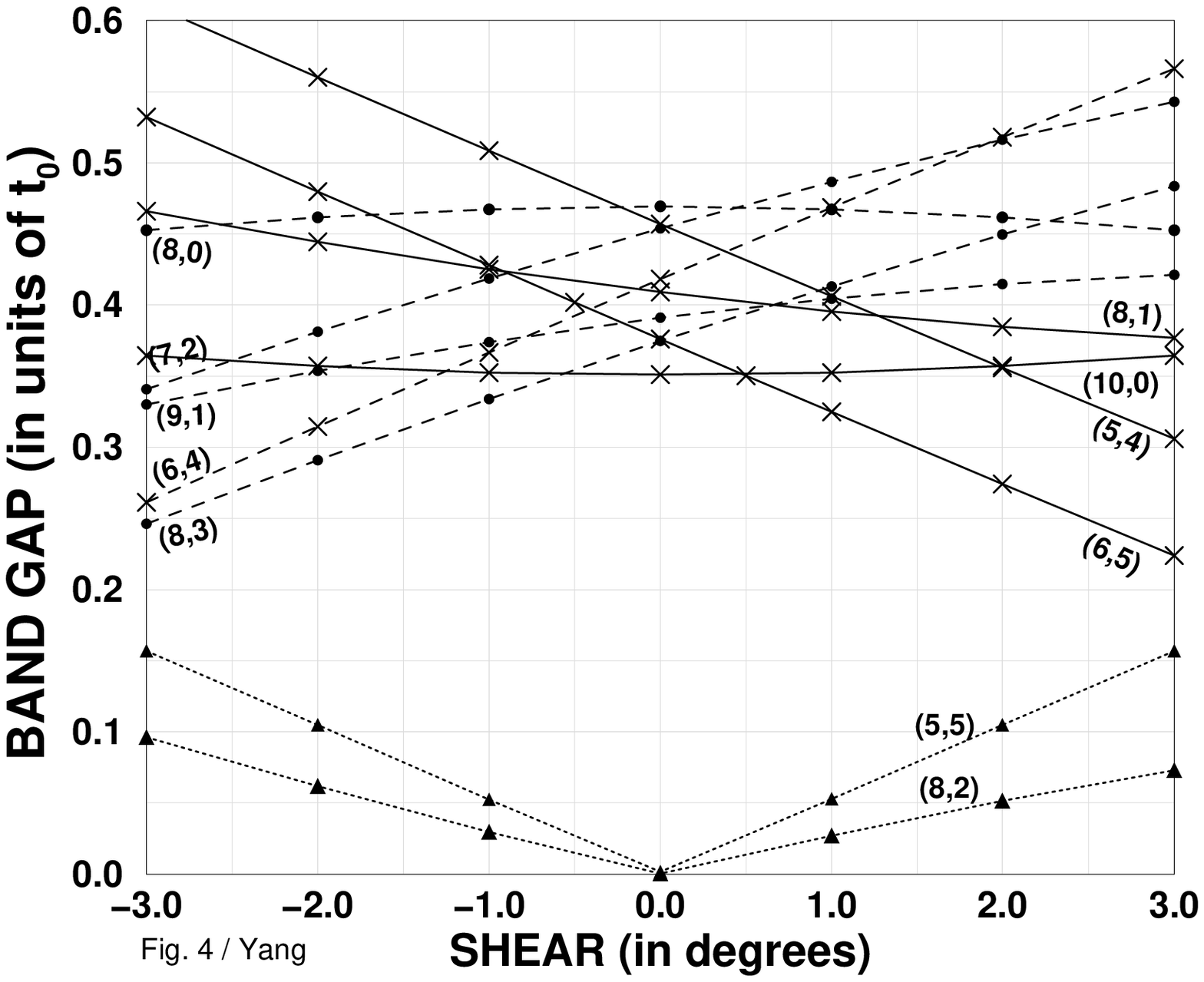,width=5in}}
\small
\end{figure}

\begin{figure}[h]
\centerline{\psfig{file=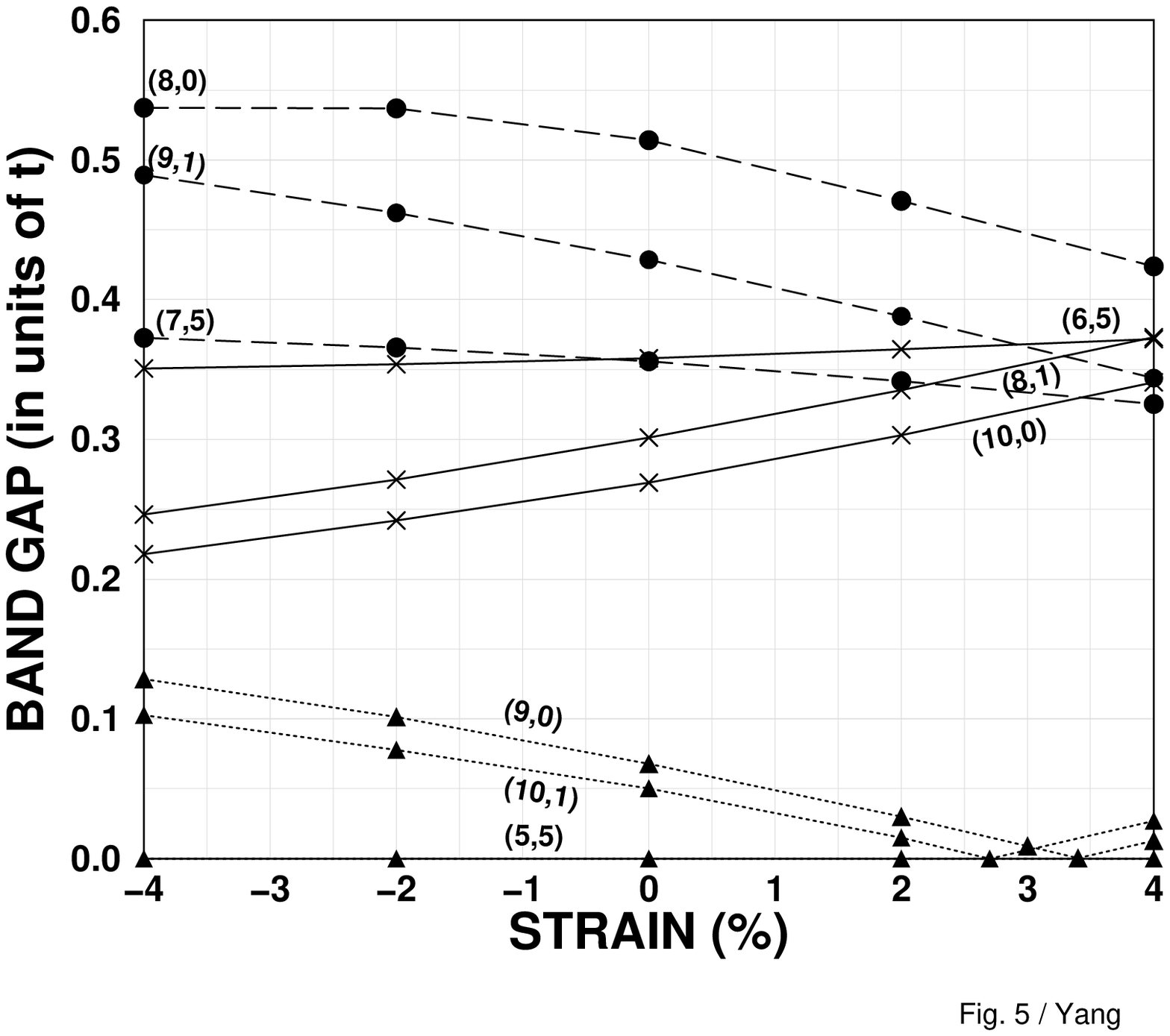,width=5in}}
\small
\end{figure}

\begin{figure}[h]
\centerline{\psfig{file=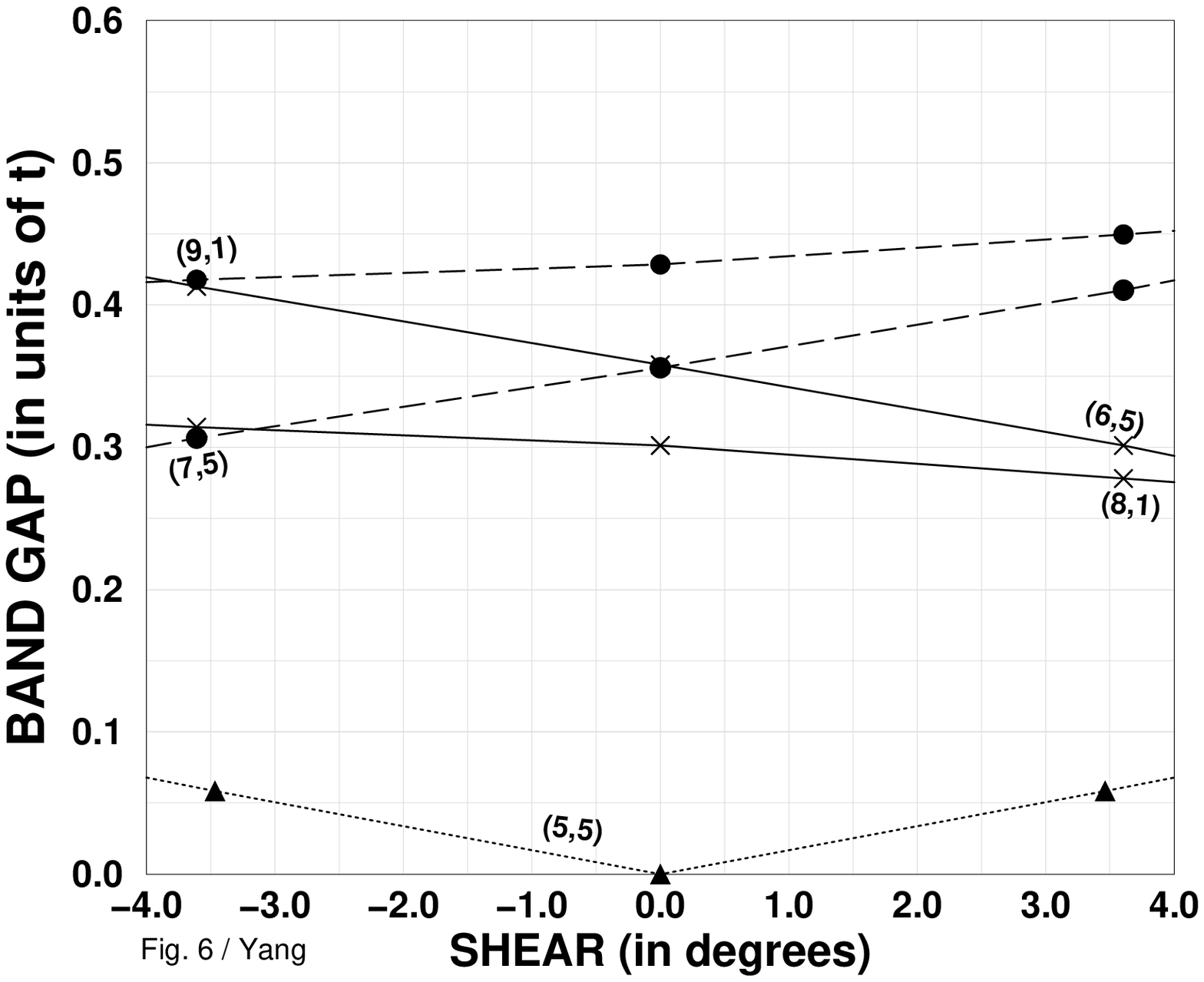,width=5in}}
\small
\end{figure}


\begin{thebibliography}{100}

\bibitem[*]{byline2} anant@nas.nasa.gov; Corresponding author

\bibitem[a]{byline4} Permanent address: CB 3255 Phillips Hall,
University of North Carolina - Chapel Hill, Chapel Hill, NC  27599;

\bibitem{Mintmire92}
J. W. Mintmire, B. I. Dunlap and C. T. White,
Phys. Rev. Lett. {\bf 68}, 631 (1992);

\bibitem{Hamada92}
N. Hamada, S. Sawada and A. Oshiyama,
Phys. Rev. Lett. {\bf 68}, 1579 (1992);

\bibitem{Saito92}
R. Saito, M. Fujita, G. Dresselhaus and M. S. Dresselhaus,
Appl. Phys. Lett. {\bf 60}, 2204 (1992)

\bibitem{Dresselhaus_book}
M. S. Dresselhaus, G. Dresselhaus and P. C. Eklund,
Chap. 19 of Science of Fullerenes and Carbon Nanotubes, Academic Press,
(1996)

\bibitem{Yakobson}
B. I. Yakobson,
App. Phys. Lett. {\bf 72}, 918 (1998);
M. B. Nardelli, B. I. Yakobson, J. Bernholc
Phys. Rev. B {\bf 57} 4277 (1998)

\bibitem{Wildoer98}
J. W. G. Wildoer,  L. C. Venema, A. G. Rinzler, R. E. Smalley and C. Dekker,
Nature {\bf 391}, 59 (1998) and
T. W. Odom, J. L. Huang, P. Kim and C. M. Lieber,
Nature {\bf 391}, 62 (1998)

\bibitem{Venema97}
L. C. Venema, J. W. G. Wildoer, H. L. J. T. Tuinstra, C. Dekker,
A. G. Rinzler and R. E. Smalley
Appl. Phys. Lett. {\bf 71}, 2629 (1997)

\bibitem{Joachim95}
C. Joachim, J. K. Gimzewski, R. R. Schlittler, and C. Chavy,
Phys. Rev. Lett. {\bf 74}, 2102 (1995)

\bibitem{Heyd97}
R. Heyd, A. Charlier, E. McRae,
Phys. Rev. B {\bf 55}, 6820 (1997).

\bibitem{Brenner_unpublished}
D. W. Brenner, Private Communication.

\bibitem{Kane97}
C. L. Kane and E. J. Mele,
Phys. Rev. Lett {\bf 78}, 1932 (1997)

\bibitem{coord_sys}
C. T. White, D. H. Robertson and J. W. Mintmire,
Phys. Rev. B {\bf 47}, 5485 (1993);
D. J. Klein, W. A. Seitz and T. G. Schmalz,
J. Phys. Chem. {\bf 97}, 1231 (1993);
P. J. Lin-Chung and A. K. Rajagopal,
J. Phys.: Condens. Matter {\bf 6}, 3697 (1994)

\bibitem{Wallace47}
This expression can be obtained by extending the treatment of
P. R. Wallace,
Phys. Rev. {\bf 71}, 622 (1947)
to the case of a graphene sheet with three bonds with unequal
bond lengths and hopping parameters.

\bibitem{Harrison}
W. A. Harrison,
Electronic structure and the properties of solids, Dover Publication Inc.,
New York (1989)

\bibitem{footnote3}
This discussion applies to the tensile case also, where
$\epsilon_{t}$ and $\epsilon_c$ should be set equal to zero. At the
end of the last paragraph we however did not use this argument to
generate the eigen spectrum because the 1D unit cell length simply
scales as $(1+\epsilon_t)$. The essence of this discussion is that
if the change in hopping parameters is accounted for, then the eigen
spectrum can be calculated by assuming that the geometry has not
changed.

\bibitem{footnote1}
defined in terms of $\{-1, 0, 1\}$; i.e., $(n-m) \bmod 3$ equal to
0, 1 and 2 corresponds to 0, 1, and -1 respectively in the notation
used.

\bibitem{aps99}
D. A. Walters, L. E. Ericson, M. J. Casavant, J. Liu, D. T. Colbert
and R. E. Smalley, "The Elastic Limit of Single-Wall Nanotube Ropes";
T. Rueckes, C. L. Cheung, J. W. Hutchinson and C. M. Lieber, 
"Tensile Strength of Carbon Nanotubes".
Both references are in 1999 centennial meeting bulletin of the American
Physical Society, Vol.44, No.1, Part II (1999), page 1818, Centennial
Meeting, American Physical, March 20-26, 1999, Atlanta, Georgia

\bibitem{Tomanek88}
D. Tomanek and S. G. Louie,
Phys. Rev. B {\bf 37}, 8327 (1988).

\bibitem{Brenner90}
D. W. Brenner
Phys. Rev. B {\bf 42}, 9458 (1990)

\bibitem{footnote2}
$2 t_0 \delta r_i$ can be replaced by $\delta t_i$ in this expression.







\end{thebibliography}
\end{document}